\documentclass[letterpaper]{article} 
\usepackage{aaai23}  
\usepackage{times}  
\usepackage{helvet}  
\usepackage{courier}  
\usepackage[hyphens]{url}  
\usepackage{graphicx} 
\usepackage{natbib}  
\usepackage{caption} 
\usepackage{algorithm}
\usepackage{algorithmic}

\usepackage{amsmath}

\usepackage{newfloat}
\usepackage{listings}
\DeclareCaptionStyle{ruled}{labelfont=normalfont,labelsep=colon,strut=off} 
\floatstyle{ruled}
\newfloat{listing}{tb}{lst}{}
\floatname{listing}{Listing}
\title{A Framework for the Construction of a Sentiment-Driven Performance Index: The Case of DAX40}
\author {
    Fabian Billert\textsuperscript{\rm 1, 2},
    Stefan Conrad\textsuperscript{\rm 1},
}
\affiliations {
    \textsuperscript{\rm 1} Heinrich-Heine University of D\"{u}sseldorf\\
    \textsuperscript{\rm 2} GET Capital AG\\
    Fabian.Billert@uni-duesseldorf.de, Stefan.Conrad@uni-duesseldorf.de
}

\begin{document}

\maketitle

\begin{abstract}
We extract the sentiment from german and english news articles on companies in the DAX40 stock market index and use it to create a sentiment-powered pendant. Comparing it to existing products which adjust their weights at pre-defined dates once per month, we show that our index is able to react more swiftly to  sentiment information mined from online news. Over the nearly $6$ years we considered, the sentiment index manages to create an annualized return of $7.51\%$ compared to the $2.13\%$ of the DAX40, while taking transaction costs into account. In this work, we present the framework we employed to develop this sentiment index.
\end{abstract}

\section{Introduction}

Advances in natural language processing in the last few years have led to an increasing amount of research on extracting information from news, social media and other sources \cite{Min.2021}. This allowed researchers to start explaining existing phenomena from a new point of view. In a financial context, one of the most promising and now more easily available features has been the investor sentiment, which expresses the opinion of investors towards different financial instruments \cite{Ferreira.2021}, \cite{Li.2018}. Recent work has found that the sentiment is an important factor for predicting the future market development (\cite{Wan.2021}, \cite{Sidogi.2021}, \cite{Ranco.2015}, \cite{Koratamaddi.2021}, \cite{Allen.2019}, \cite{Feng.2021}), although the extent of its impact is not clear. As a result, researchers are still investigating how to optimally use this new information. \\
Some works use the sentiment as a feature along with other, more conventional features such as historical returns and volatility to improve the prediction of the return of a stock on a specific day. This is often done by using complex neural networks in a time-series prediction approach (\cite{Ko.2021}, \cite{Sai.2020}, \cite{Sidogi.2021}, \cite{Sen.2021}). \\
Another branch of works uses the sentiment of different instruments in a portfolio optimization approach to find the best allocation \cite{Ferreira.2021}. Here, approaches often use reinforcement learning in a combination with neural networks to determine the optimal allocation for each day. Koratamaddi \textit{et al.} use an adaptive deep deterministic policy gradient approach and combine it with sentiment information, showing that the result outperforms the baseline without sentiment \cite{Koratamaddi.2021}. Malandri \textit{et al.} use the sentiment in combination with an LSTM for five different portfolios and show that the sentiment information consistently improves the performance \cite{Malandri.2018}. \\

In this work, we use a new approach to create a sentiment-powered performance index, using the universe of the german DAX40 as our proof-of-concept. This creates the possibility for investors to include the public opinion in their investment decisions in a convenient way. \\
The paper is organized as follows: First, we take a look at existing sentiment indices and explain their approach. We continue by giving a rough overview of the workflow in this report, from the raw text to the calculation of the sentiment index. Here, we introduce the data we used and expand upon each of the different steps in the workflow. Finally, we show the results for our sentiment index and compare it with the performance of the DAX40.

\section{Existing Products}

In recent years, the sentiment information has been used more and more to create products which take the mood of investors into account. A few companies have created automatic solutions in the form of sentiment indices, which have related ETFs or other vehicles the public can invest into.\\
At the end of 2021, S\&P Dow Jones Indices, in collaboration with Twitter, launched a set of sentiment indices based on the S\&P 500 which mine the sentiment of Tweets concerning the companies in the original index and use it to optimize the constituent's weights. The S\&P 500 Twitter Sentiment Index gathers daily sentiments and picks the 200 companies with the best sentiment once a month, weighting them by their market cap \cite{sptwitter}. Their second index, the S\&P 500 Twitter Sentiment Select Equal Weight Index, instead selects only the best 50 companies and weights them equally \cite{sptwitterequal}. The indices have performed well thus far - creating a slow but steady outperformance of the S\&P 500, although their track record is not very long. \\
A different example of an existing sentiment index is the BUZZ NextGen AI US Sentiment Leaders Index, which takes into account all large cap companies based in the US \cite {buzz}. This Index has a longer track-record, back to the end of 2015, although the related ETF has only been incepted in 2021. According to the ETF provider, the sentiment is determined using content "from online sources including social media, news articles, blog posts and other alternative datasets" \cite{buzzetf}. The 75 companies with the highest sentiment are included in the index and weighted according to their sentiment, with a maximum weight of $3\%$. This index is also adjusted once a month. \\

The downside of these products is that they are rebalanced only on a monthly basis, so the strategies cannot react to sentiment signals immediatly. Big, impactful events are usually absorbed by the markets much faster, on the timescale of days or even hours \cite{Wan.2021}. This means that the existing strategies miss out on most of the sentiment signals. In the next chapters, we investigate a different approach and try to create an improved product which is able to adapt more swiftly to opinionated events.

\section{Workflow}

In this section we expand on the different steps in our workflow. Starting with the news articles, we first preprocess them, before determining the sentiment. Since we usually have multiple articles for a single company on a given day, we then aggregate the sentiment to get a single value per company. From there, we calculate a weight value for each company using only the acquired sentiment value.

\subsection{Data}

The news article data we use consists of more than $2.85$ million articles for the forty companies in the german DAX40 as of May $2022$. The data spans the time from $01.01.2017$ to $04.09.2022$ and was scraped from german and english media websites. We received this data in a batch. To get the relevant articles for each company, we searched for articles which use the company name in their headline. For some companies, we adjust the search filter to exclude results that do not refer to the company, but a different entity. An example for this is the german sports clothing company Puma, where we found a lot of articles referring to the animal puma. For other companies, such as the automotive companies like BMW, we excluded entries reporting cases of theft or singular accidents, since small-scale events like this are unlikely to influence the stock price in any way but emit a strong sentiment. Those articles were removed by checking for the relevant keywords (like "zoo" for puma) in the headline and body of the article. After those filters, we are left with a total of around $630$k documents.

\subsection{Preprocessing}

Before preprocessing the data, we remove a subset of articles which we deemed counterproductive to a potential forecast. A few websites regularly publish an article summarizing the previous trading day, adressing the top performers in its headline. The sentiment contained in those articles obviously refers to the performance of the past, which we deemed detrimental to determining an indication for the future performance. These articles were removed by scanning the text for key phrases such as "dieser artikel wurde automatisch erstellt" - meaning "this article was created automatically". \\
We further remove duplicate articles, as well as articles with a headline longer than $1000$ tokens. Finally, we move forward using only the headlines of the articles, which are transformed to lowercase.

\subsection{Sentiment Classification}

To extract the sentiments from the headlines, we use a BERT-based model architecture. The model is based on the distilbert model in \cite{Sanh.2019} and was finetuned first with the Financial Phrasebank dataset from \cite{Malo.2014} and then again with manually annotated financial news headlines. Since the model was trained on english data, but a large number of our news articles are in german, we had to find a solution to include the german-language articles. Rather than using two separate models or retraining the model as a multilanguage model, we translated the german articles to english before classifying their sentiment. For this, we use the de-to-en translation model from \cite{Ng.2019}. To confirm that the translation does not lead to a steep decline in accuracy regarding the sentiment classification, we tested the model on around $1500$ manually annotated headlines of our dataset (for both languages). We found that for the english articles, we achieve a balanced accuracy of $67\%$, while for the german-to-english translated articles we still achieve $65\%$. \\
The model returns probabilities for three classes, negative, neutral and positive. Instead of simply taking the determined class, we use the probability and multiply it with $-1$ for the negative class and $0$ for the neutral class. This yields a value in the interval $[-1,1]$ for each news article.

\subsection{Sentiment Aggregation}

\begin{figure*}[h]
    \centering
    \includegraphics[width=0.95\textwidth]{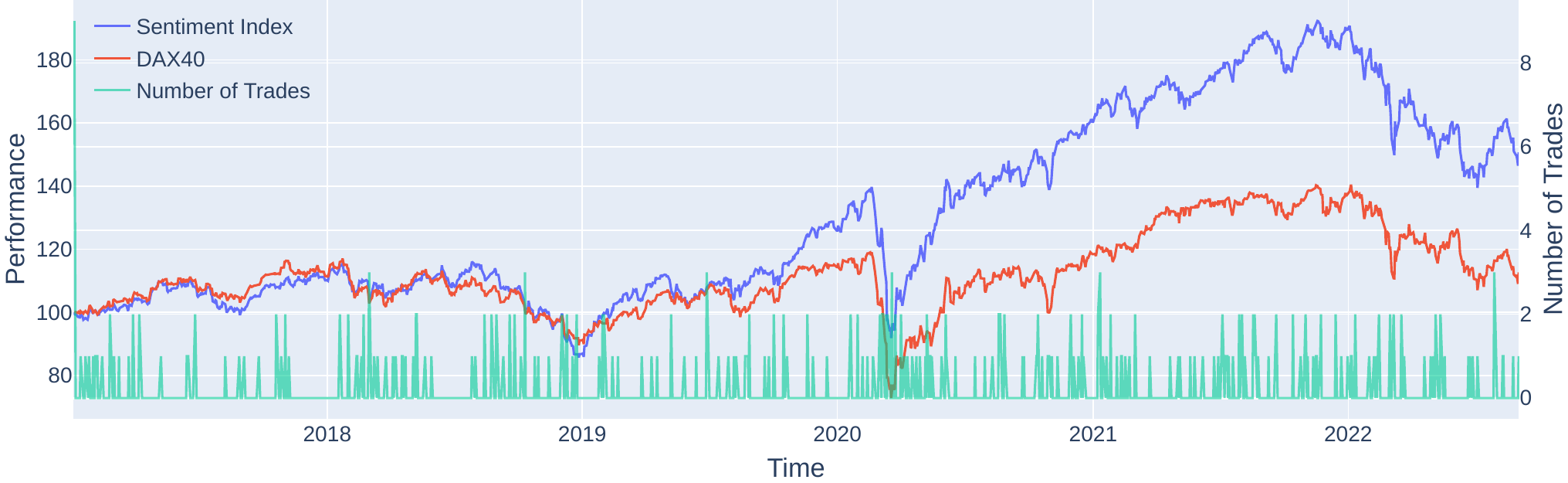}
    \caption{Left axis: Performance of the Sentiment index and the DAX40. Right axis: Number of transactions.}
    \label{fig:results}
\end{figure*}

The news articles are published at different times throughout the day. Since the german stock exchanges close at $5$:$30$ pm local time, we can not consider articles published later in the day. To make sure we do not include any forward bias, we push all articles published after $5$ pm to the next day. Following that, we calculate a single sentiment score for each day by taking the mean of all sentiment values. On days where no articles for a company are available, we set its sentiment to zero. \\
To stabilize the sentiment on days where a low amount of articles exhibits a very strong sentiment, we further apply a scaling to the sentiment values. More specifically, for a single company, we look at the amount of different sources that publish an article on a given day and multiply the sentiment with an adjustment factor that we calculate like so:
\begin{align}
    adj_i^t &=  \frac{u_i^t}{mean(u_i^{0,...,t-1})} & \text{if } u_i^t < mean(u_i^{0,...,t-1}) \\
    adj_i^t &= 1 & \text{else},
\end{align}
where $u_t$ is the number of unique sources on day $t$ for company $i$. This means that the sentiment of a company is scaled towards $0$ if the number of unique sources writing about this company is lower than the previous mean.

\subsection{Weight Determination}\label{sec:res_weights}

Using the daily sentiments for the different companies, we want to determine the individual weights that maximize the total sentiment of the universe in the next step. Obviously, this can be achieved by simply picking the company with the highest sentiment and setting its weight to one. However, due to different constraints, this is unrealistic. Firstly, different market regulations prohibit such instruments in many countries. Secondly, this would rack up huge transaction costs since $100\%$ of the index has to be churned on most days. \\
We try to solve this problem with a convex optimization approach using the python library cvxpy \cite{Agrawal.2018}, \cite{Diamond.2016}. This library allows the user to maximize an objective function under a set of constraints. The objective function maximized in our approach is as follows:
\begin{align}
    S = \sum_i w^t_is^t_i - \delta |w^{t-1}_i - w^t_i|
\end{align}
where $w^t_i$ and $s^t_i$ are the weight and sentiment of instrument $i$ at time $t$ respectively. $\delta$ is a penalty factor which reflects the transaction costs necessary to change the weights of an instrument. In classical approaches where an expected return is optimized, this factor would be equal to the transaction costs. In our case, the factor has the scale of the sentiment that needs to change for the transaction to be worth it. In order to trade only for signigifant sentiment changes, we set the penalty factor to $1$, which is one polarity. \\
As for constraints, we decide to limit the total investment in a single company to $10\%$. In addition, we require the sum of weights to be equal to one, so it is not possible to be invested in cash\footnote{In reality, the optimizer cannot fulfill such a stringent constraint, since the weights of the single stocks change every day. We instead require the sum of weights to be between $99\%$ and $99.9\%$}.

\subsection{Sentiment Index}

Having determined a way to calculate the weights for the different companies in the index, it is straightforward to determine the returns of the index:
\begin{align}
    r_{ind}^t = \sum_i w_i^t \cdot r_i^t - TC_i^t,
\end{align}
where $TC_i^t$ are the transaction costs, and $r_i^t$ is the return of stock $i$ on day $t$, which we calculate as follows:
\begin{align}
    r^t = \frac{P^t - P^{t-1}}{P^{t-1}}
\end{align}
with $P^t$ being the end-of-day price on day $t$. We set the trading costs to $0.05\%$ of the price of the stock, which is multiplied with the amount of stocks traded to get the total cost of a transaction. Consequently, this part of the equation is zero for a stock if we do not trade it. \\

\section{Results}

 In Figure \ref{fig:results}, we show the performance of the sentiment index (in blue) and compare it to the DAX40 (in red). On the second y-axis, the number of trades is depicted.
We can see that the sentiment index shows a similar performance in $2017$ and $2018$, but manages to outpace the DAX in the following three years. In the current year, the index lags slightly behind again. Especially just before and shortly after the corona-crisis, the index pulls well ahead of the DAX. In total, the index was able to generate an annualized return of $7.51\%$ compared to the $2.13\%$ of the DAX over the more than $5$ years presented here. \\
In that time, the sentiment index performs $309$ trades, $153$ of which are single trades. Single trades are usually performed when an instrument goes above the upper bound of $10\%$ we set for the optimization. Other than that, on $128$ days the index trades twice and on $18$ days it trades 3 times. Disregarding the initial investment, there are never more than $3$ transactions in a single day. In conclusion, this approach of a sentiment index performs trades more frequently, but only for cases with very strong sentiments.

\section{Conclusion}
In this work, we present a framework for determining sentiment values from news headlines and transforming them into weights for a performance sentiment index, achieving a significant outperformance compared to the benchmark index over the five year time span we observed. We compare our approach to existing products, which have pre-defined re-balancing dates in contrast to our more flexible approach which is able to change the constituent weights at any point in time without amassing large amounts of transaction costs. \\
We apply the framework to the german stock market, the DAX40, as an example. However, it is straightforward to apply it to any other universe of stocks. \\
In future works, we would like to adapt this framework for portfolios, where it is possible to reduce the investment ratio and hold cash. This opens a new dimension as the general market sentiment is important, as opposed to just the sentiment of the constituents. In addition, we plan on continually improving different areas of the framework and compare new results to the here presented baseline approach. \\
One such area is the sentiment model, which could be refined further, for example by using a larger amount of annotated data to finetune it. In addition, the use of the penalty factor can be investigated and refined further, for example by using different values for different market situations.

\section{Acknowledgements}

This work was partially funded by GET Capital AG. The authors would further like to thank pressrelations GmbH for the support in infrastructure they provided.

\bibliography{sentiment_index_Final}

\end{document}